\documentclass[preprintnumbers,amsmath,amssymb,showpacs,floatfix,11pt,prd,onecolumn,superscriptaddress,nofootinbib]{revtex4}
\usepackage{graphicx}
\usepackage{epsfig}
\usepackage{bm}
\usepackage{amsfonts}
\begin{document}

\title{Noether Gauge Symmetry of Modified Teleparallel Gravity Minimally Coupled with a Canonical Scalar Field }

\author{\textbf{Adnan Aslam}}
\email{adnangrewal@yahoo.com}
\affiliation{Center for Advanced Mathematics and Physics (CAMP),
National University of Sciences and Technology (NUST), H-12,
Islamabad, Pakistan}

 \author{\textbf{Mubasher Jamil}}\email{mjamil@camp.nust.edu.pk}
\affiliation{Center for Advanced Mathematics and Physics (CAMP),
National University of Sciences and Technology (NUST), H-12,
Islamabad, Pakistan}\affiliation{Eurasian International Center
for Theoretical Physics, Eurasian National University, Astana
010008, Kazakhstan}

 \author{\textbf{ Davood Momeni}}\email{d.momeni@yahoo.com}
 \affiliation{Eurasian International Center
for Theoretical Physics, Eurasian National University, Astana
010008, Kazakhstan}

 \author{\textbf{Ratbay Myrzakulov}}\email{
rmyrzakulov@csufresno.edu}\affiliation{Eurasian International Center
for Theoretical Physics, Eurasian National University, Astana
010008, Kazakhstan}

\begin{abstract}
\textbf{Abstract:} This paper is devoted to the study of  Noether
gauge symmetries of $f(T)$ gravity minimally coupled with a
canonical scalar field. We explicitly determine the unknown
functions of the theory $f(T),V(\phi), W(\phi)$. We have shown that
there are two invariants for this model, one of which defines the
Hamiltonian $H$ under time invariance (energy conservation) and the
other is related to scaling invariance. We show that the equation of
state parameter in the present model can cross the cosmological
constant boundary. The behavior of Hubble parameter in our model
closely matches to that of $\Lambda$CDM model, thus our model is an
alternative to the later.
\end{abstract}

\maketitle
\newpage
\section{Introduction}

Observational data from Ia supernovae (SNIa) show that currently the
observable Universe is in accelerated expansion phase \cite{riess}.
This cosmic acceleration has also been confirmed by different
observations of large scale structure (LSS) \cite{3} and
measurements of the cosmic microwave background (CMB) anisotropy
\cite{4}. The essence of this cosmic acceleration backs to ``dark
energy'',  an exotic energy which generates a large negative
pressure, whose energy density  dominates the Universe (for a review
see e.g. \cite{sami}). The astrophysical nature of dark energy
confirms that it is not composed of baryonic matter. Now from
cosmological observations we know that the Universe is spatially
flat and consists of about 70\% dark energy, 30\% dust matter (cold
dark matter plus baryons) and negligible radiation. But the nature
of dark energy as well as its cosmological origin remains mysterious
at present.

One of the methods for constructing  a dark energy model is to
modify the geometrical part of the Einstein equations. The general
paradigm consists in adding into the effective action, physically
motivated higher-order curvature invariants and non-minimally
coupled scalar fields. But if we relax the Riemannian manifold,  we
can construct the models based on the torsion $T$ instead of the
curvature. The representative models based on this strategy are
termed `modified gravity' and include $f(R)$ gravity \cite{fr},
Horava-Lifshitz gravity \cite{hl,hl2,hl3}, scalar-tensor gravity
\cite{st,st1} and the braneworld model \cite{brane,brane1} and the
newly $f(T)$ gravity \cite{f(T)}.

The $f(T)$ theory of gravity is a meticulous class of modified
theories of gravity. This theory can be obtained by replacing the
torsion scalar $T$ in teleparallel gravity \cite{TP} with an
arbitrary function $f(T)$. The dynamical equations of motion can be
obtained by varying the Lagrangian with respect to the vierbein
(tetrad) basis. Wu et al have proposed viable forms of $f(T)$ that
can satisfy both cosmological and local gravity constraints
\cite{puxunwu}. Further Capozziello et al  discussed the
cosmographical method for reconstruction of $f(T)$ models
\cite{Capozziello}. But the $f(T)$ model has some theoretical
problems. For example it's not locally Lorentz invariance, possesses
extra degrees of freedom, violate the first law of thermodynamics
and inconsistent Hamiltonian formalism \cite{sotiriou}.

In the past, the use of scalar fields in certain physical theories,
especially particle physics, has been explored. This led to study
the role of scalar fields in cosmology as well. Recently some of the
present authors investigated the behavior of scalar fields in $f(T)$
cosmology. In \cite{epjc1}, we introduced a non-minimally
conformally coupled scalar field and dark matter in $f(T)$
cosmology. We investigated the stability and phase space behavior of
the parameters of the scalar field by choosing an exponential
potential and cosmologically viable form of $f(T)$. We found that
the dynamical system of equations admit two unstable critical
points; thus no attractor solution existed in that model. In another
investigation \cite{epjc2}, we studied the Noether symmetries (which
are symmetries of the Lagrangian) of $f(T)$ involving matter and
dark energy. In that model, the dark energy was considered as a
canonical scalar field with a potential. The analysis showed that
$f(T)\sim T^{3/4}$ and $V(\phi)\sim \phi^2$.  Therefore it becomes
meaningful to reconstruct a scalar potential $V(\phi)$ in the
framework of $f(T)$ gravity. It was demonstrated that dark energy
driven by scalar field, decays to cold dark matter in the late
accelerated Universe and this phenomenon yields a solution to the
cosmic coincidence problem \cite{jamil}. In this paper, we study the
Noether gauge symmetries (NGS) of the model, which provide a more
general notion of the Noether symmetry. This approach is useful in
obtaining physically viable choices of $f(T)$, and has been
previously used for the $f(R)$ gravity and generalized
Saez-Ballester scalar field model as well \cite{jami}.

The plan of this paper is as follows: In Section II, we present the
formal framework of the $f(T)$ action minimally coupled with a
scalar field. In section III, we construct the governing
differential equations from the Noether condition and solve them in
an accompanying subsection. In section IV, we study the dynamics of
the present model. Finally we conclude this work. In all later
sections, we choose units $c=16\pi G=1$.

\section{$f(T)$ gravity}

If we limit ourselves to the validity equivalence principle, we must
work with a gauge theory for gravity and such a gauge theory is
possible only on curved manifold. Construction of a gauge theory on
Riemannian manifolds is only one option and may be the simplest one.
But it's possible to write a gauge theory for gravity, with  metric,
non-metricity and torsion can be constructed easily \cite{smalley}.
Such theories are defined on a Weitzenb\"{o}ck spacetime, with
globally non zero torsion but with vanishing local Riemannian
tensor. In this theory,  which is called teleparallel gravity,
people are working on a non-Riemannian spacetime manifold. The
dynamics of the metric is defined uniquely by  the torsion $T$. The
basic quantities in teleparallel or the natural extension of it,
namely $f(T)$ gravity, is the vierbein (tetrad) basis  $e^{i}_{\mu}$
\cite{saridakis}.  This basis of vectors  is unique and orthonormal
and is defined by the following equation
\begin{eqnarray}\nonumber
g_{\mu\nu}=e_{\mu}^{a}e_{\nu}^b \eta_{ab},\ \ {a,b}=0,1,2,3
\end{eqnarray}
This tetrade basis must be orthonormal and $\eta_{ab}$ is the flat Minkowski  metric, $e^{a}_{\mu}e_{b}^\mu =\delta^a_{b}$.
One suitable form of the action for $f(T)$ gravity in Weitzenb\"{o}ck
spacetime is \cite{ata}
\begin {equation}\label{a-1}
\mathcal{S}=\int d^{4}xe\Big((T+f(T))+\mathcal{L}_{m}\Big)\,,
\end{equation}
where
 $f(T)$ is an arbitrary function of torsion $T$ and $e=\det(e^{i}_{\mu})$. The dynamical quantity of the model is the scalar torsion $T$ and the matter Lagrangian $\mathcal{L}_m$.
The equation of motion derived from the action, by varying with respect to the $e^{i}_{\mu}$,
is given by
\begin{eqnarray}\nonumber
e^{-1}\partial_{\mu}(e S^{\:\:\:\mu
\nu}_{i})(1+f_T)-e_i^{\:\lambda}T_{\:\:\:\mu
\lambda}^{\rho}S^{\:\:\:\nu \mu}_{\rho}f_T +S^{\:\:\:\mu
\nu}_{i}\partial_{\mu}(T)f_{TT}-\frac{1}{4}e_{\:i}^{\nu}
(1+f(T))=4 \pi  e_i^{\:\rho}T_{\rho}^{\:\:\nu}\,.
\end{eqnarray}
As usual $T_{\mu\nu}$ is the energy-momentum tensor for matter
sector of  the Lagrangian $\mathcal{L}_m$. It is a straightforward
calculation to show that this equation of motion is reduced to
Einstein gravity when $f(T)=0$. Indeed, this is the equivalence
between the teleparallel theory and the Einstein gravity.

\section{Our model}

We take
a spatially flat homogeneous and isotropic
Friedmann-Lema\^{i}tre-Robertson-Walker (FLRW)
spacetime
\begin{equation}\label{1}
ds^2=-dt^2+a^2(t)[dr^2+r^2(d\theta^2+\sin^2\theta d\varphi^2)]\,.
\end{equation}
We add in the action (\ref{a-1}) a  scalar field with an unknown
potential function $V=V(\phi)$ (sometimes also known as
Saez-Ballester model \cite{sb}). However a slight redefinition of
($\Phi(\phi)=\int\limits^\phi d\phi\sqrt{\pm V(\phi)},$ where $+/-$
correspond to non-phantom/phantom phase, respectively
\cite{odintsov}). The total action reads:
\begin{eqnarray}
\mathcal{S}=\int d^4x~ e \Big(T+f(T)+\lambda (T+6H^2)+V(\phi)\phi_{;\mu}\phi^{;\mu}-W(\phi)\Big).\label{action}
\end{eqnarray}
Here trace of the torsion tensor is $T=-6H^2$, $e=det(e^\mu_i)$,
$\lambda$ is the Lagrange multiplier and $H=\frac{\dot{a}}{a}$  is
the Hubble parameter. For our convenience, we keep the original
Saez-Ballester scalar field in our effective action. Varying
(\ref{action}) with respect to $T$, we obtain
\begin{eqnarray}\label{landa}
\lambda=-(1+f'(T)).
\end{eqnarray}
Here prime denotes the derivative with respect $T$. By substituting
(\ref{landa}) in (\ref{action}), and integrating over the
spatial volume we get the following reduced Lagrangian:
\begin{eqnarray}\label{lag}
\mathcal{L}(a,\phi,T,\dot{a},\dot{\phi})=
a^3\Big[T+f(T)-[1+f'(T)]\Big(T+6(\frac{\dot{a}}{a})^2\Big)+V(\phi)\dot{\phi}^2-W(\phi)\Big].
\end{eqnarray}
For the Lagrangian (\ref{lag}), the equations of motion read as follows
\begin{equation}
f_{TT}(T+6H^2)=0, \label{6}
\end{equation}
\begin{equation}
 \frac{\ddot a}{a}=-\frac{1}{4(1+f_T)}\Big[f-Tf_T-\frac{T}{3}(1+f_T)+V(\phi)\dot\phi^2-W(\phi)+4H\dot T f_{TT}  \Big],\\
\end{equation}
\begin{equation}
\ddot \phi+3H\dot\phi+\frac{1}{2V}( V'\dot\phi^2+W')=0.
\end{equation}
Equation (\ref{6}) indicates two possibilities: (1) $f_{TT}=0$,
which gives the teleparallel gravity and we are not interested in
this case. (2) Another possibility is $T=-6 H^2$ which is the
standard definition of the torsion scalar in $f(T)$ gravity. In the
next section we will investigate the Noether gauge symmetries of the
newly  proposed model in (\ref{lag}).

\section{Noether gauge symmetry of the model}
To calculate the Noether symmetries, we define it first. A vector
field \cite{jami}
\begin{eqnarray}
\mathbf{X}=\mathcal{T}(t,a,T,\phi)\frac{\partial }{\partial
t}+\alpha(t,a,T,\phi)\frac{\partial }{\partial
a}+\beta(t,a,T,\phi)\frac{\partial }{\partial
T}+\gamma(t,a,T,\phi)\frac{\partial }{\partial \phi},
\end{eqnarray}
is a Noether gauge symmetry corresponding to a Lagrangian
$\mathcal{L}(t,a,T,\phi,\dot{a},\dot{T},\dot{\phi})$ if
\begin{equation}
\mathbf{X}^{[1]}\mathcal{L}+\mathcal{L}D_{t}(\mathcal{T})=D_{t}B, \label{Noether}
\end{equation}
holds, where $\mathbf{X}^{[1]}$ is the first prolongation of the
generator $\mathbf{X}$, $B(t,a,T,\phi)$ is a gauge function and
$D_{t}$ is the total derivative operator
\begin{equation}
D_{t}=\frac{\partial }{\partial t}+\dot{a}\frac{\partial }{\partial
a}+\dot{T}\frac{\partial }{\partial T}+\dot{\phi}\frac{\partial
}{\partial \phi}.
\end{equation}
The prolonged vector field is given by
\begin{equation}
\mathbf{X}^{[1]}=\mathbf{X}+\alpha_{t}\frac{\partial }{\partial
\dot{t}}+\beta_{t}\frac{\partial }{\partial
\dot{T}}+\gamma_{t}\frac{\partial }{\partial \dot{\phi}},
\end{equation}
where
\begin{align}
\alpha_{t}=D_{t}\alpha-\dot{a}D_{t}\mathcal{T}, ~~~  \beta_{t}=D_{t}\beta-\dot{T}D_{t}\mathcal{T},\ \
\gamma_{t}=D_{t}\gamma-\dot{\phi}D_{t}\mathcal{T}.
\end{align}
If $\mathbf{X}$ is the Noether symmetry corresponding to the
Lagrangian $\mathcal{L}(t,a,T,\phi,\dot{a},\dot{T},\dot{\phi})$, then
\begin{equation}
\mathbf{I}=\mathcal{T}\mathcal{L}+(\alpha-\mathcal{T}\dot{a})\frac{\partial
\mathcal{L}}{\partial \dot{a}}+(\beta-\mathcal{T}\dot{T})\frac{\partial
\mathcal{L}}{\partial \dot{T}}+(\gamma-\mathcal{T}\dot{\phi})\frac{\partial
\mathcal{L}}{\partial \dot{\phi}}-B, \label{firstint}
\end{equation}
is a first integral or an invariant or a conserved quantity
associated with $\mathbf{X}$.
 The Noether condition (\ref{Noether}) results in the
over-determined system of equations
\begin{align}
&\mathcal{T}_{a}=0,~~\mathcal{T}_{\phi}=0,~~\mathcal{T}_{T}=0,~~\alpha_{T}=0,   \\
&{\gamma} _{T}=0,~~B_{T}=0,~~2a^{3}V\gamma_{t}=B_{\phi},\\
&6(1+f')\alpha_{\phi}-a^{2}V\gamma_{a}=0,\\
&12a(1+f')\alpha_{t}+B_{a}=0,\\
&3V\alpha+aV'\gamma+2aV\gamma_{\phi}-aV\mathcal{T}_{t}=0,\\
&(1+f')(\alpha+2a\alpha_{a}-a\mathcal{T}_{t})+af''\beta=0,\\
&3a^{2}(f-Tf'-W)\alpha-a^{3}Tf''\beta-a^{3}W'\gamma+a^{3}(f-Tf'-W)\mathcal{T}_{t}=B_{t}.
\end{align}

We  obtain the solution of the above system of linear partial
differential equations for $f(T)$, $V(\phi)$, $W(\phi)$,
$\mathcal{T}$, $\alpha$, $\beta $ and $\gamma$. We have:
\begin{align}
f(T)&=\frac{1}{2}t_{0}T^{2}-T+c_2,\label{sol1}\\
V(\phi)&=V_{0}\phi^{-4},\label{V}\\
W(\phi)&=W_0\phi^{-4}+c_2,\label{W}\\
\mathcal{T}&=t+c_1,\\
\alpha &=a,\\
\beta &=-2T,\\
\gamma &=\phi,\label{sol2}
\end{align}
where $t_{0}, V_{0}, W_{0}, c_2$ and $c_{1}$ are constants. It is
interesting to note that quadratic $f(T)=T^2$ has been used to model
static wormholes in $f(T)$ gravity \cite{lobo}. Also the scalar
potential is proportional to $\phi^{-4}$ which has been previously
reported in \cite{jami} for $f(R)$-tachyon model. Recently Iorio \&
Saridakis \cite{sari} studied solar system constraints on the model
(\ref{sol1}) and found the bound $|t_0|\leq 1.8 \times 10^4 m^2.$
Further a $T^2$ term can cure all the four types of the finite-time
future singularities in $f(T)$ gravity, similar to that in $F(R)$
gravity \cite{bamba}. The quadratic correction to teleparallel model
is quite vital as a next approximation in astrophysical context.
Hence the Noether gauge symmetry approach generates a cosmologically
viable model of $f(T)$ gravity.

It is clear from (\ref{sol1})-(\ref{sol2}) that the Lagrangian
(\ref{lag}) admits two Noether symmetry generators
\begin{align}
\mathbf{X}_{1}&=\frac{\partial}{\partial t},\label{gen1}\\
\mathbf{X}_{2}&=t\frac{\partial}{\partial
t}+a\frac{\partial}{\partial a}-2T\frac{\partial}{\partial
T}+\phi\frac{\partial}{\partial \phi}.\label{gen2}
\end{align}
The first symmetry $\mathbf{X}_{1}$ (invariance under time
translation) gives the energy conservation of the dynamical system
in the form of (\ref{inter1}) below, while the second symmetry
$\mathbf{X}_{2}$ (scaling symmetry) and a corresponding conserved
quantity of the form (\ref{inter2 }) below. The two first integrals
(conserved quantities) which are
\begin{align}
\mathbf{I}_{1}=-\frac{1}{2}&t_{0}T^{2}a^{3}+6t_{0}Ta\dot{a}^{2}-V_{0}a^{3}{\phi}^{-4}\dot{\phi}^{2}-W_{0}a^{3}\phi^{-4},\label{inter1}\\
\mathbf{I}_{2}=-\frac{1}{2}&t_{0}tT^{2}a^{3}+6t_{0}tTa\dot{a}^{2}-V_{0}ta^{3}{\phi}^{-4}\dot{\phi}^{2}-W_{0}ta^{3}\phi^{-4}\nonumber\\
&-12t_{0}Ta^{2}\dot{a}+2V_{0}a^{3}\phi^{-3}\dot{\phi}.\label{inter2
}
\end{align}
Also the commutator of generators satisfies $[\mathbf{X}_1,\mathbf{X}_2]=\mathbf{X}_1$
which shows that the algebra of generators is closed.

\section{Cosmological implications}
We rewrite Eq. (\ref{lag}) using the solutions for $f(T)$,
$V(\phi)$, $W(\phi)$ as obtained in the previous section in the
following form
\begin{eqnarray}
\mathcal{L}= \frac{\dot{a}^4}{a}+\frac{a^3(V_0
\dot{\phi}^2-W_0-c_2\phi^4)}{\phi^4},
\end{eqnarray}
where $t_0$ is an arbitrary constant. We choose $t_0=\frac{1}{18}$
for simplification and further we take $c_2\neq0$.  The
Euler-Lagrange equations read
\begin{equation}
\frac{\ddot a}{a}=\frac{H^2}{4}+\frac{1}{4a\phi^4 H^2}\Big(V_0 \dot\phi^2-W_0-c_2\phi^4\Big)\label{a},\\
\end{equation}
\begin{equation}
\ddot\phi+3H\dot\phi-\frac{2\dot
\phi^2}{\phi}=\frac{2W_0}{V_0\phi}.\label{phi}
\end{equation}
Their evolutionary behavior is obtained by numerically solving the Euler-Lagrange  equations (\ref{a})-(\ref{phi}) for an appropriate set of the parameters and the initial conditions.
\begin{figure*}[thbp]
\begin{tabular}{rl}
\includegraphics[width=7.5cm]{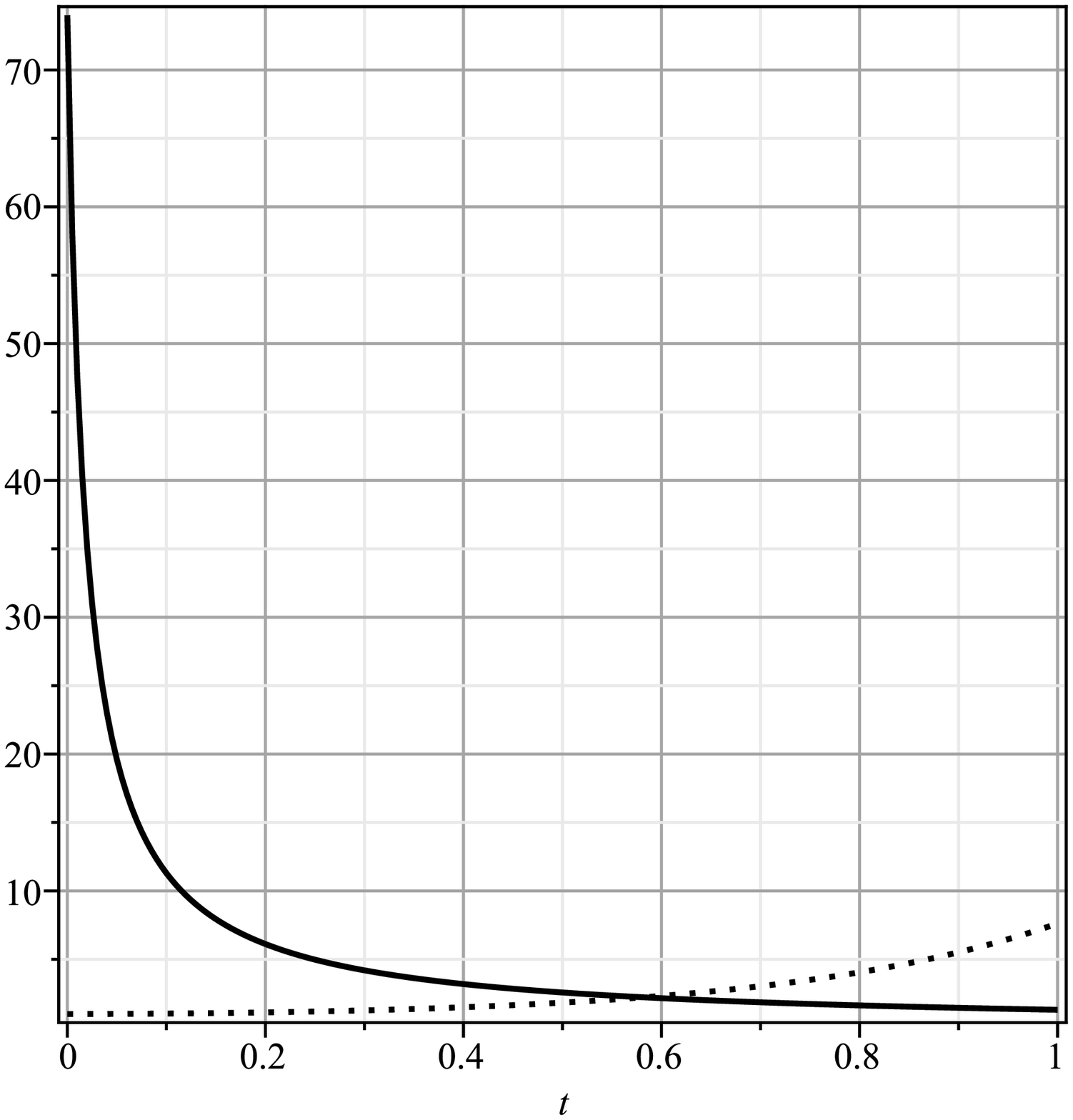}&
\includegraphics[width=7.5cm]{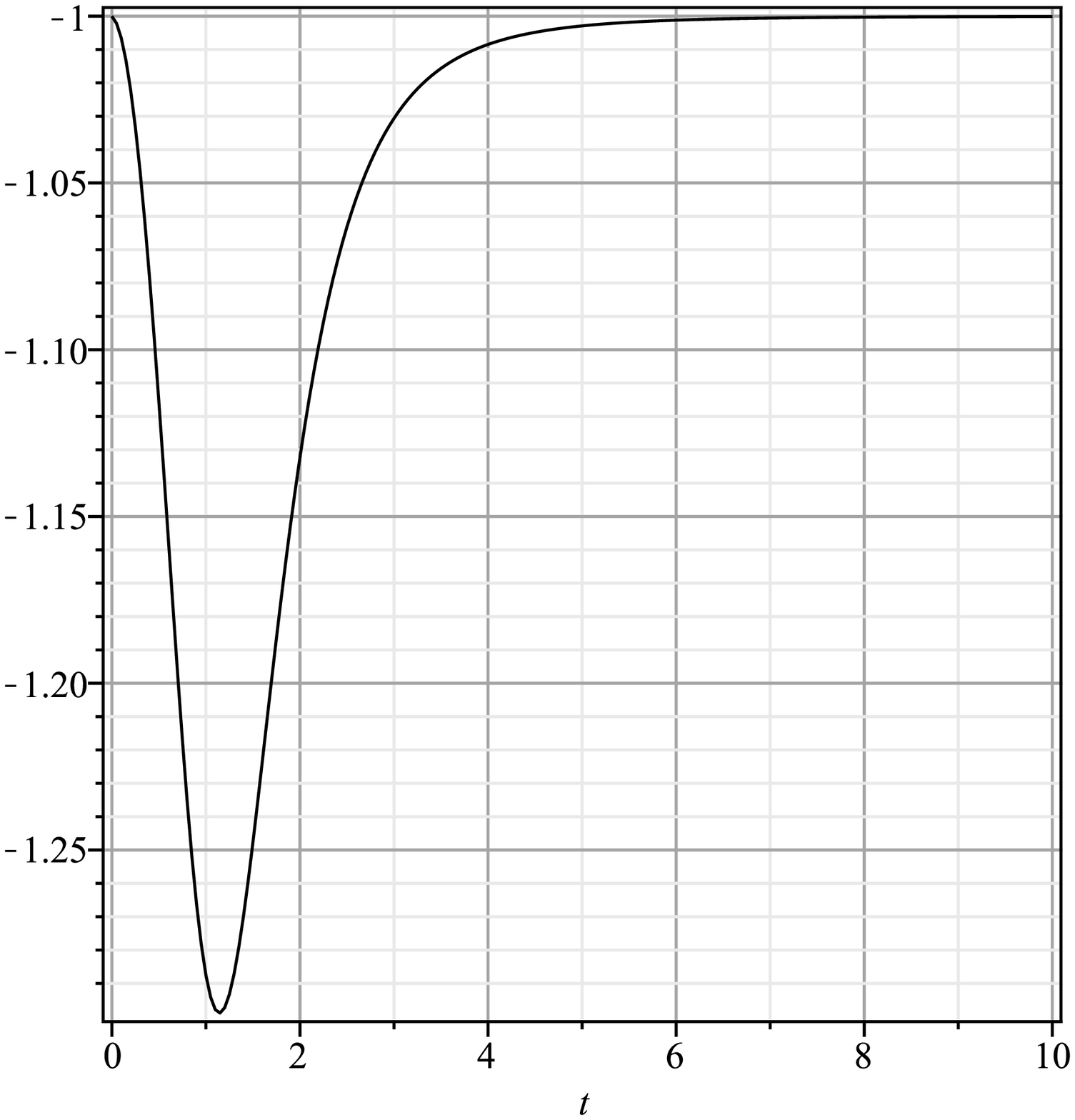} \\
\end{tabular}
\caption{ (\textit{Left})  Cosmological evolution of $H(t),\phi(t)$
vs time $t$. The model parameters chosen as $V_0=1,$ $W_0=2 V_0$.
Various curves correspond to (solid, $H(t)$), (dots, $\phi(t)$).
(\textit{Right})  Variation of $w$ vs time $t$. The model parameters
are chosen as $\alpha_0=2,$ $\beta=\frac{\rho_{m0}}{V_0}$,
$\omega=1$. In both figures, we chose the initial conditions $\dot
a(0)=H_0$, $\phi(0)=0.1$,  $ a(0)=1$, $\dot \phi(0)=1$. }
\end{figure*}
To obtain the equation of state parameter numerically,  first we
note that for $f(T)$-Saez-Ballester theory, the energy-momentum
tensor reads
\begin{equation}
T_{\mu\nu}=V(\phi)\Big[g_{\mu\nu}\phi_{;\alpha}\phi^{;\alpha}-2\phi_{;\mu}\phi_{;\nu}\Big]-g_{\mu\nu}W(\phi).
\end{equation}
It is easy to calculate the total energy density and averaged pressure
\begin{eqnarray}
\rho&=&V(\phi)\dot\phi^2-W(\phi),\\
p&=&-V(\phi)\dot\phi^2-W(\phi).
\end{eqnarray}
The EoS parameter is constructed via
\begin{eqnarray}\label{w}
w\equiv\frac{p}{\rho}=\frac{V(\phi)\dot\phi^2+W(\phi)}{-V(\phi)\dot\phi^2+W(\phi)}.
\end{eqnarray}
Putting the potential functions (\ref{V}) and (\ref{W}) in (\ref{w}), we get
\begin{equation}
w=\frac{\dot\phi^2+\beta\phi^4+\alpha_0}{-\dot\phi^2+\beta\phi^4+\alpha_0},\quad
\beta=\frac{c_2}{V_0},\quad\alpha_0=\frac{W_0}{V_0}.
\end{equation}
The numerical simulation of $w,\dot w$ is drawn in the figure which
shows that $w$ behaves like the phantom energy for a brief period of
time. This conclusion is exciting since there exists convincing
astrophysical evidence that the observable Universe is currently in
the phantom phase \cite{caldwell}.

\section{Conclusion}

In this paper, motivated by some earlier works on Noether symmetry
in $f(T)$ gravity, we introduced a new model containing a canonical
scalar field with a potential. Firstly we showed that this model
obeys a quadratic term of torsion, with potential proportional to
$\phi^{-4}$ which also appears for tachyonic field in $f(R)$ model.
It is also interesting to note that quadratic $f(T)=T^2$ has been
used to model static wormholes in $f(T)$ gravity in literature.

We mention our key results and comments below:
\begin{itemize}

\item Our numerical simulations show that there happens a phantom
crossing scenario for a brief period in this toy model, after which
the state parameter evolves to cosmological constant asymptotically.

\item  The behavior of Hubble parameter in our model closely
mimics to that of $\Lambda$CDM model, thus our model is an
alternative to the later.

\item Since at the same time $\Lambda$CDM model
 cannot explain the phantom crossing as is observed from the
empirical astrophysical results, one should prefer alternatives to
$\Lambda$CDM model such as the present one. It is curious to note
that such a result is obtained from a Lorentz invariance violating
$f(T)$ theory, however, the same theory is consistent with solar
system tests, contains attractor solutions, and is free of massive
gravitons.

\item The results reported here are significantly different from
\cite{epjc1,epjc2} since there we calculated the `Noether
symmetries' while here only `Noether gauge symmetries' are obtained.
As one can see, the results obtained here are different from the
ones previously obtained in the literature.

\item The most important feature of $f(T)$ gravity which differs
it from any other ``curvature invariant'' model like $f(R)$ theory,
is its irreducibility to a scalar model in the Jordan frame unlike
$f(R)$. As we know, $f(R)$ gravity can be reduced to a scalar field
by a simple identification between the scalar field and the gravity
sector of the action in the Jordan frame. But here, since the
Lorentz symmetry is broken locally, and also for leakage of finding
such formal transformation between the scalar field and the Torsion
action, these two models are not equivalent. So unless the $f(R)$,
here proposition of the scalar field is not artificial and do not
add any additional degree of freedom to the model. From another
point of the view, the $f(T)$ gravity is not conformal invariant, so
reduction of the gravity sector to the scalar matter is not
possible. So indeed by introducing the scalar field in the action,
we avoided from a similar formal extensions like two scalar
components models, like Quintessence. Note that proposition of the
scalar field to $f(T)$ is completely new irreducible action but to
$f(R)$ is just the quintom model or in it's extreme form reduces to
the multi-scalar models with less symmetry than the original action.

\end{itemize}

\subsection*{Acknowledgment}

We would gratefully thank the anonymous referee for useful criticism
on our paper.

\end{document}